\documentclass[manuscript]{aastex}
\def\simge{\mathrel{%
    \rlap{\raise 0.511ex \hbox{$>$}}{\lower 0.511ex \hbox{$\sim$}}}}
\def\simle{\mathrel{
    \rlap{\raise 0.511ex \hbox{$<$}}{\lower 0.511ex \hbox{$\sim$}}}}

\slugcomment{submitted to Ap. J. Lett.}
\shorttitle{Photochemistry of Hot Jupiter Stratospheres}
\shortauthors{Zahnle et al.}

\begin{document}
\title{Atmospheric Sulfur Photochemistry on Hot Jupiters}

\author{K.\ Zahnle}
\affil{NASA Ames Research Center, Moffett Field, CA 94035}
\email{Kevin.J.Zahnle@NASA.gov}

\author{M.\ S.\ Marley}
\affil{NASA Ames Research Center, Moffett Field, CA 94035}
\email{Mark.S.Marley@NASA.gov}

\author{R.\ S.\ Freedman}
\affil{NASA Ames Research Center, Moffett Field, CA 94035}
\email{freedman@darkstar.arc.nasa.gov}

\author{K.\ Lodders}
\affil{Washington University - St. Louis}

\and
\author{J.\ J.\ Fortney}
\affil{Department of Astronomy and Astrophysics, University of California - Santa Cruz}

%\documentclass[11pt]{article}
%usepackage{geometry}                % See geometry.pdf to learn the layout options. There are lots.
%\geometry{letterpaper}                   % ... or a4paper or a5paper or ...
%\geometry{landscape}                % Activate for for rotated page geometry
%\usepackage[parfill]{parskip}    % Activate to begin paragraphs with an empty line rather than an indent
%\usepackage{graphicx}
%\usepackage{amssymb}
%\usepackage{longtable}
%\usepackage{natbib}
%\usepackage{booktabs}
%\usepackage{epstopdf}
\DeclareGraphicsRule{.tif}{png}{.png}{`convert #1 `dirname #1`/
`basename #1 .tif`.png}

\begin{abstract}
% \noindent {\bf Abstract}

We develop a new 1D photochemical kinetics code to address stratospheric chemistry and stratospheric heating in hot Jupiters.
Here we address optically active S-containing species and CO$_2$ at $1200 \leq T\leq 2000$ K.
HS (mercapto) and S$_2$ are highly reactive species that are generated photochemically and thermochemically from H$_2$S 
with peak abundances between 1-10 mbar. 
S$_2$ absorbs UV between 240 and 340 nm and is optically thick for metallicities $[{\rm S}/{\rm H}]>0$ at $T\geq 1200$ K. 
HS is probably more important than S$_2$, 
as it is generally more abundant than S$_2$ under hot Jupiter conditions and it absorbs at somewhat redder wavelengths.    
We use molecular theory to compute an HS absorption spectrum from sparse available data and find that HS should absorb strongly between 300 and 460 nm, 
with absorption at the longer wavelengths being temperature sensitive. 
When the two absorbers are combined, radiative heating (per kg of gas) peaks at 100 $\mu$bars,
with a total stratospheric heating of $\sim\!8\times 10^4$ W/m$^2$ for a jovian planet orbiting a solar-twin at 0.032 AU. 
Total heating is insensitive to metallicity.
The CO$_2$ mixing ratio is a well-behaved quadratic function of metallicity,
 ranging from $1.6\times10^{-8}$ to $1.6\times 10^{-4}$ for $-0.3<[{\rm M}/{\rm H}]<1.7$.
CO$_2$ is insensitive to insolation, vertical mixing, temperature ($1200<T<2000$), and gravity.
The photochemical calculations confirm that CO$_2$ should prove a useful probe of planetary metallicity.
\end{abstract}


\section{Introduction}

Stratospheric temperature inversions are ubiquitous in the Solar System, and it is beginning to look as if they are commonplace on 
hot Jupiters as well.
Stratospheric temperature inversions form when substantial amounts of light are
absorbed at low pressures (high altitudes) where radiative cooling is inefficient.  
Hubeny et al.\ (2003) pointed out that efficient absorption of visible light by gaseous TiO and VO would
greatly heat the upper atmospheres of those planets already hot enough for these molecules to be present as vapor.
Thermal inversions on transiting hot Jupiters were first seen by Richardson et al.\ (2007) for HD 209458b and Harrington et al.\  (2007) for HD 149026b.
The observed flux ratio at 8 $\mu$m for HD 149026b agreed only with models that included a thermal inversion (Fortney et al.\ 2006).
Temperature inversions have since been confirmed by \emph{Spitzer}
observations of HD 209458b (Knutson et al.\ 2008a), XO-1b (Machalek et al.\ 2008),
and TrES-4 (Knutson et al.\ 2009), all of which show distinctive flux  
ratios in IRAC bands that suggest inversions (Fortney et al.\ 2006; Burrows et al.\ 2007).   
More circumstantial evidence exists for HD 179949b (Barnes et al.\ 2008).

On the other hand TrES-1, the least irradiated planet with published \emph{Spitzer}
observations, does not appear to have a pronounced inversion (Burrows et al.\ 2008).  Nor,
seemingly, does HD 189733b, which is also modestly irradated  (Charbonneau et al.\
2008; Barman et al.\ 2008). 
%One suggestion %that addresses why some planets have hot stratospheres, and others not,
One suggestion is that temperature inversions are triggered by irradiation reaching a
critical level that is hot enough to evaporate TiO and VO from grains,
as discussed by Burrows et al.\ (2007), Fortney et al.\ (2008), and Burrows et al.\ (2008).
%Cooler atmospheres that lack TiO and VO would not have stratospheres.
However, irradiation of XO-1b and HD 189733b is within uncertainties the same (Torres et al.\ 2008), which poses a challenge to the irradiation trigger.

In the Solar System, stratospheric temperature inversions are often caused by  
absorption of UV light by gases or aerosols produced by photochemistry.  
Here we ask if atmospheric chemistry might play a similar role in hot Jupiters. 
Speculation has tended to focus on sulfur-containing species (Tinetti 2008),
%(``polysulphur,'' Tinetti 2008, god knows how to refer to a video of a conference presentation), 
as the reservoir species H$_2$S is expected to be abundant (Visscher et al 2006) in these atmospheres
and many of its breakdown products (S$_2$, in particular) absorb violet and ultraviolet light.    

\section{The Photochemical Model}

Previous photochemical modeling of hot Jupiters addressed the abundance of
photochemical H (Liang et al 2003) and the absence of photochemical smogs (Liang et al 2004).
Liang et al (2003) focused on the high H/H$_2$ ratio that arises from H$_2$O photolysis.
In their second paper, Liang et al (2004) argued that simple hydrocarbons would not condense
to form photochemical smogs in hot solar composition atmospheres.
Neither study considered sulfur. 

We have developed a new general purpose 1D
photochemical kinetics code applicable to hot extrasolar planets.
The code is based on the sulfur photochemistry model for early Earth
originally described by Kasting et al (1989) and Kasting (1990),
and subsequently adapted by Zahnle et al (2006) and Claire et al
(2006) to address sulfur photochemistry of Earth's
atmosphere during the Archean, and by Zahnle et al.\ (2008) to
address martian atmospheric chemistry.
Steady state solutions are found by integrating the system through
time using a fully implicit backward-difference method.

Our chemical network has been upgraded from that used by Zahnle et al (1995) to address the chemistry
generated when the fragments of Comet Shoemaker Levy 9 struck Jupiter. 
We have assembled a reasonably complete list of the reactions that
can take place between the small molecules and free radicals that can be made from H, C, O, N, and S.
The code solves 507 chemical reactions for 49 chemical species:
H, H$_2$O, OH, O, O$_2$, CO, CO$_2$, HCO, H$_2$CO, C, CH, CH$_2$, CH$_3$,
CH$_4$, CH$_3$O, C$_2$, C$_2$H, C$_2$H$_2$, C$_2$H$_3$, C$_2$H$_4$, C
$_2$H$_5$, C$_2$H$_6$,  C$_4$H, C$_4$H$_2$, CN,  HCN, N, N$_2$, NO,
NH, NH$_2$, NH$_3$, NS, H$_2$S, HS, S, S$_2$, S$_3$,  S$_4$, S$_8$,
SO, HSO,
SO$_2$, OCS,  CS, HCS, H$_2$CS, CS$_2$, and H$_2$.
Reaction rates, when known, are selected from the publicly available
NIST database (http://kinetics.nist.gov/kinetics).
In order of decreasing priority, we choose between reported reaction rates
according to relevant temperature range, newest review, newest experiment, and newest theory.
Reverse reaction rates $k_r=K_{\rm eq}k_f$ of two-body reactions are determined from the forward reaction rate $k_f$ and the
equilibrium $K_{\rm eq}=\exp{\left\{\left(-\Delta H + T\Delta S\right)/RT\right\}}$ by using $H^{\circ}(T)$ and $S^{\circ}(T)$ as available ($R$ is the gas constant).  
Rates are not available for all reactions, especially for reactions involving elemental sulfur.
%Space limitations preclude presenting detailed reaction rate tables here.
We will present a full listing of the chemical reactions important to sulfur in a 
more general followup study.

Here we use simple descriptions of atmospheric properties.
The background atmosphere is 84\% H$_2$ and 16\% He.
We include Rayleigh scattering by H$_2$ (Dalgarno and Williams 1962).
For our base case we assume an isothermal atmosphere with T=1400 K;
constant vertical eddy diffusivity $K_{zz}=1\times 10^{7}$ cm$^2$/s;
a surface gravity of 20 m/s$^2$; and insolation
levels $I$ by a solar twin that are 1000$\times$ greater than at Earth.
Metallicity proved to be the most interesting parameter and was varied $-0.3 \leq [{\rm M}/{\rm H}] \leq 1.7$.
In these units, solar metallicity is $[{\rm M}/{\rm H}]=0$, Jupiter's is $[{\rm M}/{\rm H}]=0.5$, and Saturn's is $[{\rm M}/{\rm H}]=0.8$.
Short chemical lifetimes of S-containing species make our results insensitive to $K_{zz}$.
Model parameters are listed in Table 1.

At the upper boundary we set a zero flux lid at 1 $\mu$bar, with neither escape nor exogenous supply.
For the lower boundary we use fixed equilibrium mixing ratios of the most abundant species
at 1 bar of H$_2$ and temperature $T$ (Lodders and Fegley 2002, Visscher et al 2006).
For other species we force the mixing ratio at 1 bar to approach zero. 
We scale the lower boundary conditions such that
the total mixing ratios of C, O, N, and S all scale linearly with metallicity.

Absorption by S$_2$ between 240 nm and $\sim$360 nm from the ground
state is analogous to the Schumann-Runge system in O$_2$ (Okabe 1978).
Strong, distinctive S$_2$ emission near 300 nm was observed on Jupiter  
after the impact with Shoemaker-Levy 9 with Jupiter in 1994 (Noll et
al 1995).  Subsequent thermochemical modeling showed that S$_2$  
readily forms as a major product in a shock-heated
($T>1000$ K) gas of either cometary or jovian composition (Zahnle et al 1995, Zahnle 1996).  
S$_2$ has also been seen in gases vented by volcanoes on Io (Spencer et al, 2000; Moses et al 2002).
For S$_2$, we use absorption cross sections at 1500 K
computed by van der Heijden and van der Mullen (2001).

\begin{figure}[!htb]  
\centering
\includegraphics[width=1.0\textwidth]{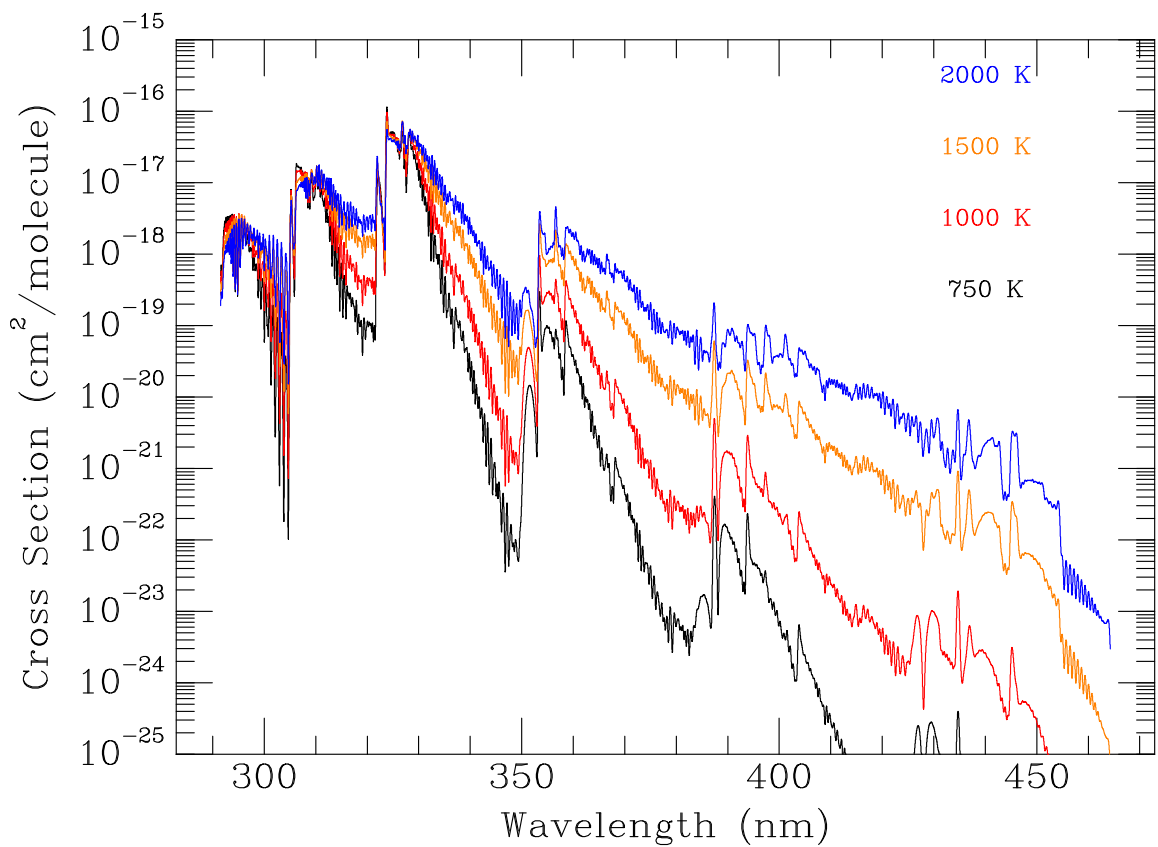} 
    \caption{Theoretical absorption cross sections of HS radicals at near UV, violet and indigo wavelengths at four temperatures at 30 mbar pressure.
Cross sections were computed from the lowest five vibrational levels of the ground electronic state $X^2\Pi$ to the lowest three vibrational levels of the upper level $A^2\Sigma^+$.
The excited vibrational levels of $A^2\Sigma^+$ are strongly predissociating, which suggests that absorption at wavelengths shorter than 324 nm is probably continuous rather
than allocated into the well-defined bands shown here. }
\label{fig:figure_SH}
\end{figure}

The HS (mercapto) radical absorbs from its ground state at 324 nm (Okabe 1978).
Visscher et al (2006) predicted that HS would be very abundant in equilibrium at hot Jupiter conditions.
We find the same.   
We therefore calculated absorption cross sections of HS 
at four temperatures at 30 mbar pressure
 using literature values of the molecular properties. 
 The ground $X^2\Pi$ state has been well studied (Ram et al 1995) but the upper level $A^2\Sigma^+$
is subject to strong pre-dissociation
(Resende and Ornellas 2001, Wheeler et al 1997, Schneider et al 1990, Henneker and Popkie 1971), and only the value of the rotational constant $B$
 and the spacing of the lowest vibrational energy levels have been well measured. 
 Using these constants and a value for the electronic band oscillator strength of the 0-0 transition derived
from a study of HS in the solar spectra by Berdyugina and Livingston (2002),
 a line list was computed using the RLS code developed by R.N.\ Zare and D.\ Albritton (Zare et al 1973).
This RLS code uses the molecular constants and band strengths to predict line positions and strengths by fitting to an RKR potential (Zare et al 1973). 
Other needed data  --- Franck-Condon factors, partition functions, etc.\ --- were derived either from the cited literature,
the program itself, or from Sauval and Tatum (1984) or Larsson (1983).
The calculations were carried out for values of $v^{\prime\prime}$(0-4) % and v' [012] 
and $v^{\prime} (012)$.
Because the excited vibrational levels of the $A^2\Sigma^+$ state are unstable with respect to predissociation, 
the corresponding optical transitions are likely to be broad and shallow, or even continuous. 
These uncertainties principally affect the absorption spectrum at wavelengths shorter than 324 nm, which is in the range that is absorbed strongly by S$_2$.
Results are shown in Figure \ref{fig:figure_SH}. 
In the photochemical model we used only the 1500 K absorption coefficients.

Other sulfur allotropes are better absorbers than S$_2$ but less abundant.
S$_3$ absorbs strongly between 350 and 500 nm, and S$_4$ absorbs between 450 nm and 600 nm, but more weakly (Billmers and Smith 1991).
Unfortunately, the chemistries of S$_3$ and S$_4$ are very uncertain, and we have had to estimate the important reaction rates. 
In an earlier version of this study, we focused on the heats of formation, and we tentatively concluded that S$_3$ heating would be important for metallicities $[{\rm S}/{\rm H}]>0.7$. 
We have since learned that reactions of the form H + S$_n$ $\rightarrow$ HS + S$_{n-1}$, where $n\geq 2$, are strongly favored by entropy.
The revised model predicts less S$_2$ and much less S$_3$, which reduces the importance of S$_3$ heating considerably.

Sulfanes (H$_2$S$_n$, hydropolysulfides) will be present in cooler hot Jupiters. 
At low temperatures sulfanes absorb VUV between 260 nm and 330 nm (Steudel and Eckert 2003).
Absorption may extend beyond 400 nm at higher temperatures as the ground state becomes vibrationally excited, 
as in HS, but to first approximation these wavelengths are covered by the more abundant S$_2$ and HS.
We have not included sulfanes in this study.

  \section{Results}

\begin{figure}[!htb] %  figure placement: here, top, bottom
   \centering
\includegraphics[width=1.0\textwidth]{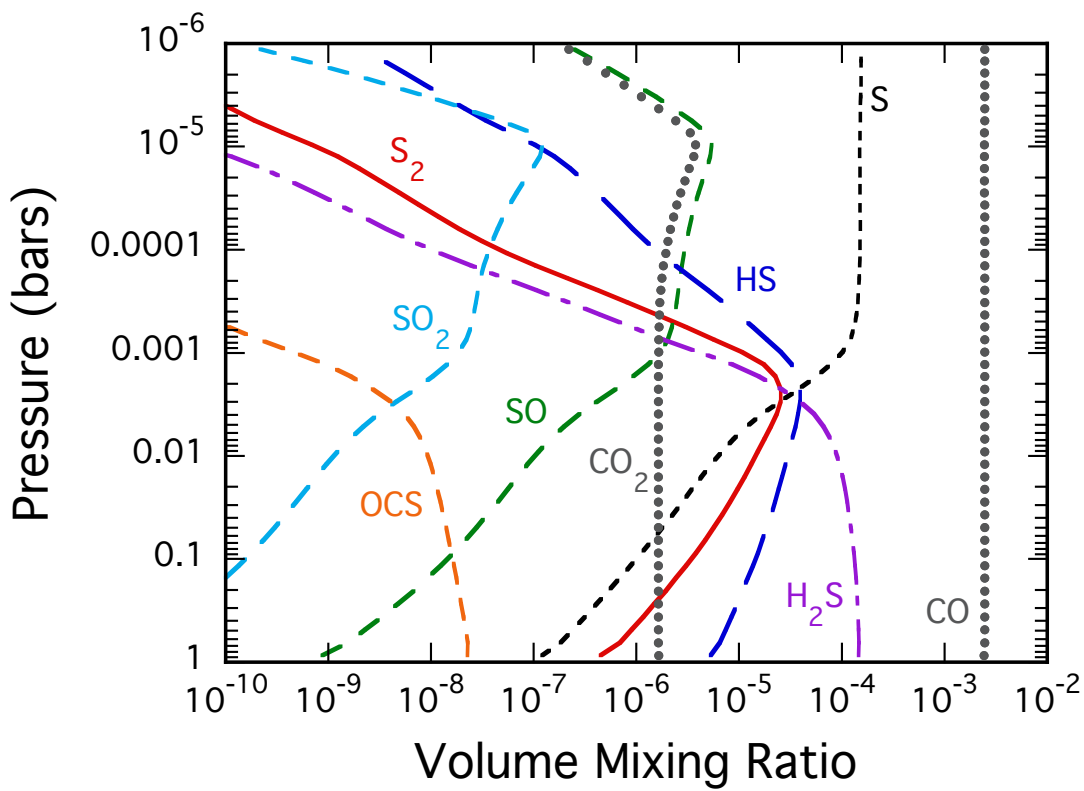} % requires the graphicx package
   \caption{\small Important sulfur species, CO, and CO$_2$ in the atmosphere of a hot Jupiter with a ``planetary'' metallicity of $[{\rm M}/{\rm H}]=0.7$.  
 The atmosphere is assumed isothermal at 1400 K and insolated 1000$\times$ more strongly than Earth.  
Other model M parameters are listed in Table 1. 
 The prominent transition at $\sim$2 mbar --- where the S$_2$ mixing ratio peaks --- is associated with photolysis of H$_2$S.
 % H$_2$S, CO and HS scale linearly with metallicity, CO because it is the major C-bearing species.
 %S$_2$, SO, CO$_2$, and OCS, which scale as metallicity squared, and SO$_2$, which scales as metallicity cubed, are all much more abundant than in the solar composition atmosphere of Figure \ref{fig:model_A}.
 The bump in CO$_2$ at 6 $\mu$bars is attributable to photochemistry.
Abundance profiles in the 1400 K atmosphere are generally representative of atmospheres with $1200\leq T \leq 2000$ K.}
\label{fig:model_M}
\end{figure}

Figure \ref{fig:model_M} shows how CO$_2$ and the abundant S-containing species vary as a function of altitude.
 This particular case shows a hot Jupiter at 1400 K with a ``planetary'' metallicity of $[{\rm M}/{\rm H}]=0.7$.
Figure \ref{fig:model_M} is broadly representative of all our models with $1200 \leq T \leq 2000$ K and $-0.3 < [{\rm M}/{\rm H}] < 1.7$. 
In particular, S$_2$ and HS show well-defined peaks at $\sim\!2$ mbars that coincide with the altitude where H$_2$S photolysis becomes important.
At lower altitudes H$_2$S is the main S-containing species, and at higher altitudes S is.  It is
also notable that the atmosphere becomes more oxidizing at higher
altitudes where H$_2$O photolysis is important.

Table 1 lists some key results pertinent to sulfur for several variations of basic model parameters.
The models assume that $K_{zz} = 10^{7}$ cm$^2$/s and $g=2000$ cm/s$^2$ unless otherwise noted.
In this temperature range the models are insensitive to $K_{zz}$ (results not shown).
Model G shows that, as expected, column densities vary inversely with $g$.
 
Column densities of S$_2$ and HS are sensitive to metallicity. 
To first approximation, species with one metal atom, such as H$_2$O and H$_2$S, increase linearly with metallicity, and
species with two metal atoms, such as SO and S$_2$, increase as the square of metallicity (VIsscher et al 2006).
% and species with three metal atoms, such as SO$_2$, increase as the cube. 
A slight complication is that CO and N$_2$ increase linearly with metallicity because these are the major reservoir species for C and N, respectively; hence
CO$_2$ increases as the square of metallicity (as CO $\times$ O), rather than as the cube.   

The models are not sensitive to temperature and insolation over the parameter ranges ($1200 \leq T \leq 2000$ K and $1\leq I \leq 1000$) presented here.
Insensitivity of the chemistry to $T$ and $I$ surprised us, and suggests that thermochemical equilibrium is more important for sulfur than photochemistry or kinetics.
Minor differences are that HS is favored by higher temperatures and SO and S$_2$ are favored by high $I$.  
Not shown here is that the chemistry changes markedly for $T < 1100$ K:  hydrocarbons, CS, and CS$_2$ become abundant, and the results become sensitive to $K_{zz}$. 
Cooler atmospheres introduce a variety of new topics best left for another study.
   
Carbon dioxide, a robust molecule and a potential observable, has been reported in HD 189733b by Swain et al (2009).
CO$_2$ is generated from CO by reaction with OH radicals.
The chief source of OH is the reaction of H$_2$O with atomic hydrogen;
 at high altitudes UV photolysis of H$_2$O is also important.
%There is a strong tendency for the CO$_2$/CO ratio to approach thermochemical equilibrium.
%The pertinent chemical reactions are well-characterized, % and the CO and H$_2$ abundances are fixed by metallicity,
%so the computed CO$_2$ mixing ratio is robust and in good agreement with the predictions of equilibrium thermochemistry.
We find that CO$_2$ mixing ratios range from $1.6\times10^{-8}$ to
$1.6\times 10^{-4}$ for $-0.3\leq [M/H] \leq 1.7$, scaling as the square of metallicity.
Table 1 lists computed CO$_2$ mixing ratios in the models discussed here.
These results are insensitive to insolation, vertical mixing, temperature between 1200 K and 2000 K, and gravity.
The CO$_2$/CO ratio is nearly independent of pressure, as seen in Figure \ref{fig:model_M}.
Pressure independence is expected because the controlling reactions, CO$_2$+H $\leftrightarrow$ CO + OH and  H+H$_2$O $\leftrightarrow$ H$_2$ + OH,
and the controlling equilibrium, CO$_2$+H$_2$ $\leftrightarrow$ CO + H$_2$O, all leave the total pressure unchanged.
(At very high altitudes photochemistry alters the CO$_2$/CO ratio.)
The computed CO$_2$ abundances are in good agreement with the
reported observation of CO$_2$ at the ppmv level in HD 189733b (Swain et al. 2009).
The sensitivity of CO$_2$ to metallicity and insensitivity to other atmospheric parameters makes CO$_2$ a good probe of planetary metallicity,
as pointed out by Lodders and Fegley (2002).

\subsection{Optical depth and stratospheric heating}

\begin{figure}[!htb] %  figure placement: here, top, bottom
   \centering
\includegraphics[width=1.1\textwidth]{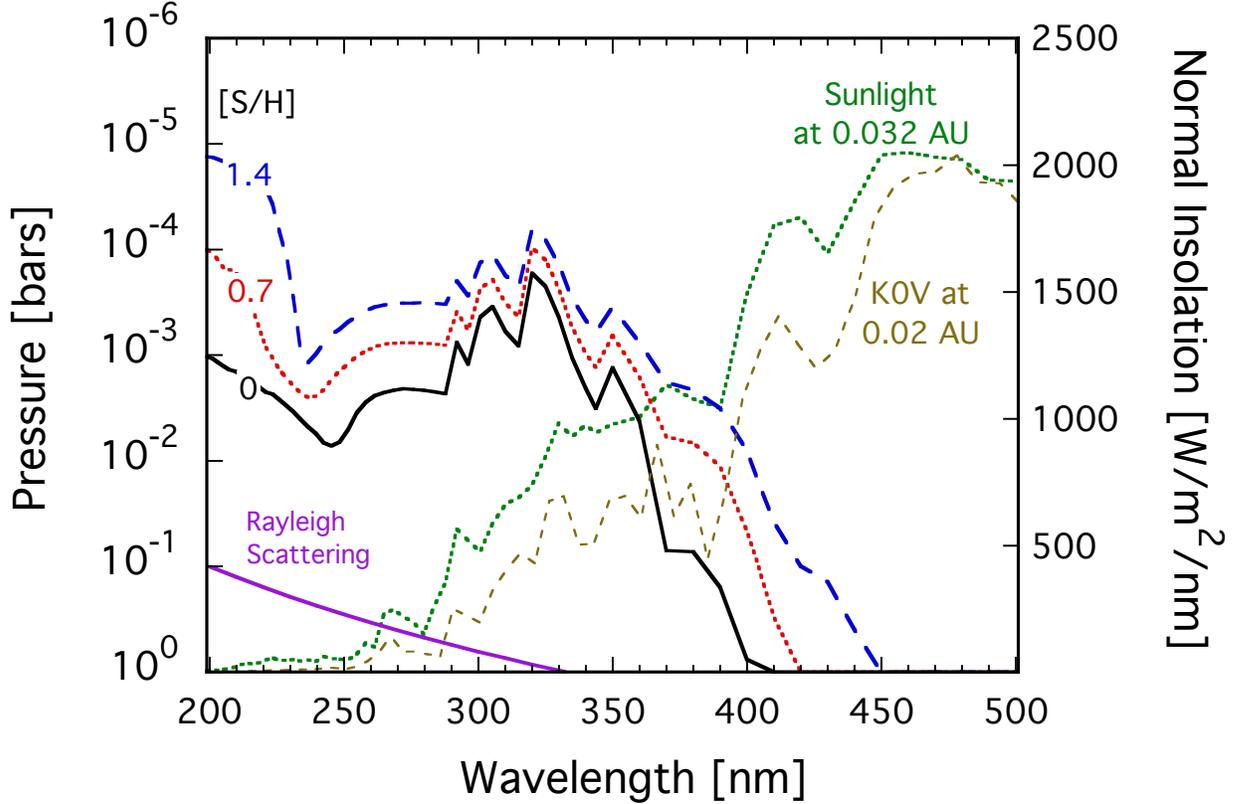} % requires the graphicx package
   \caption{Pressure levels of the $\tau=1$ surface as a function of wavelength for three metallicities, $[{\rm S}/{\rm H}]=0$, 0.7, and 1.4.
 These metallicities correspond to models A, M, and MM of Table \ref{table_one}. 
Absorption between 250 and 300 nm is mostly by S$_2$ and absorption between 300 and 460 nm is by HS.
Structure blueward of 324 nm is associated with transitions to predissociating states and is probably fictitious.
 The $\tau=1$ surface of a pure H$_2$ Rayleigh scattering atmosphere and two incident stellar spectra, one for the Sun and
 another for a generic K0V dwarf, are shown for comparison.
}
\label{fig:figure_UV}
\end{figure}

Figure \ref{fig:figure_UV} shows the pressure levels where the solar and
planetary metallicity atmospheres of Models A, M, and MM become
optically thick.  Opacity is dominated by HS, with some contribution by S$_2$ at wavelengths shorter than 300 nm.
The twin peaks between 300 nm and 320 nm may be fictitious, but the peak at 324 nm could prove diagnostic of HS.
A solar and a K0V stellar spectrum are shown for comparison.

\begin{figure}[!htb] 
   \centering
\includegraphics[width=1.0\textwidth]{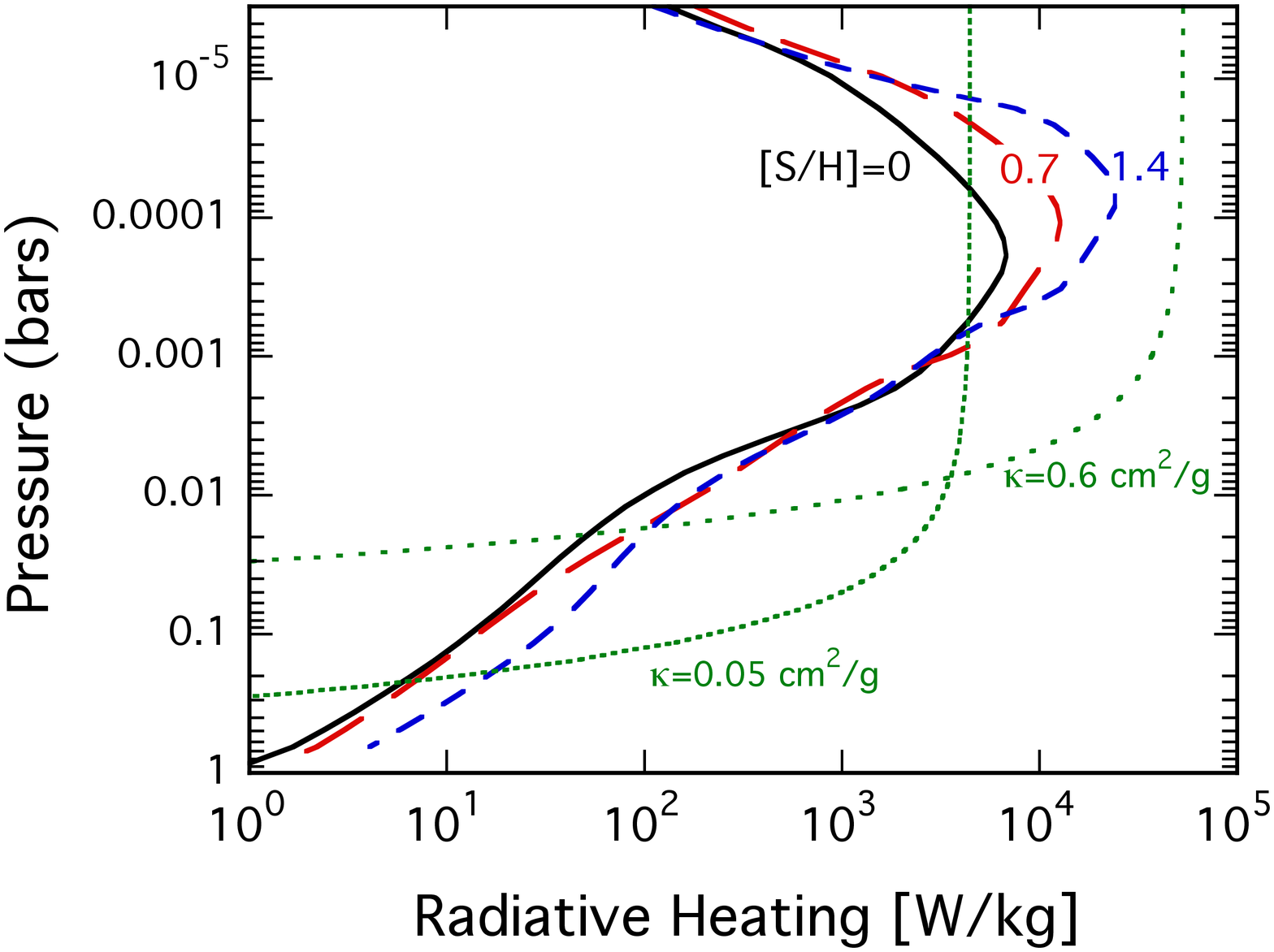} 
   \caption{Radiative heating at different altitudes for three metallicities, $[{\rm S}/{\rm H}]=0$, 0.7, and 1.4.
   These correspond to models A, M, and MM of Table \ref{table_one}. 
   Heating rates are given in W/kg, which emphasizes the potential impact on temperature. 
   Heating peaks at 100 $\mu$bars but extends through the stratosphere.
   Heating with constant gray opacities of $0.05$ and $0.6$ cm$^2$/g for $430 < \lambda < 1000$ nm is shown for comparison.
}
\label{fig:figure_heat}
\end{figure}

Figure \ref{fig:figure_heat} shows the magnitude of stratospheric heating
and the pressure level where the heating occurs for a solar-twin primary at 0.032 AU ($I=1000$)
for 3 metallicities (Models A, M, and MM).
Radiative heating is dominated by HS, and is nearly saturated through the stratosphere for all these models
(see also Table 1).
By contrast, peak heating at $\sim\! 100\, \mu$bars takes place where SO and SO$_2$ are significant.
The sensitivity of SO and SO$_2$ to metallicity is reflected in greater heating rates at $\sim\! 100\, \mu$bars.

Cumulative stratospheric heating rates for these models are listed in Table 1. 
For a solar-twin at 0.032 AU, cumulative heating above 1 mbar is typically $4\times 10^{4}$ W/m$^2$ 
and above 0.1 bars is typically $8\times 10^{4}$ W/m$^2$, i.e., about half the energy is absorbed
in the lower stratosphere. 
Burrows et al. (2008) modeled hot stratospheres by adding an unknown gray absorber.
They found that gray cross-sections of $0.05-0.6$ cm$^2$/g, averaged over 430 to 1000 nm for altitudes above 0.03 bars,
could produce the observed heating.
Heating profiles using gray opacities in this range are
plotted for comparison on Fig \ref{fig:figure_heat} for the same planet and star.
The gray opacities produce more heating in total (indeed, the stratospheres in both these models are optically thick), 
and more heating at low altitudes, but at higher altitudes sulfur generates heating at levels quite similar to what Burrows et al find useful.

\section{Conclusions}

We develop a new 1D photochemical model for 
stratospheric modeling of hydrogen-rich atmospheres of warm or hot exoplanets.
This model is applicable to any H-rich planet subject to high insolation,
including hot Neptunes, superearths, and waterworlds.
Here we apply the model to sulfur chemistry, stratospheric heating, and CO$_2$ abundance.

We find that hot stratospheres of hot Jupiters could be explained by absorption of UV and violet visible light by HS and S$_2$, two highly reactive species that
are generated chemically from H$_2$S.
For a hot Jupiter orbiting a solar-twin at 0.032 AU, for a wide range of possible planetary compositions,
HS and S$_2$ together absorb $4\times 10^{4}$ W/m$^2$ at altitudes above 1 mbar and another $4\times 10^{4}$ W/m$^2$ at altitudes between
 1 mbar and 0.1 bar.
 This level of heating approaches what Fortney et al (2006) and Burrows et al (2008) use in their
most successful LTE spectral models.
Non-LTE mechanisms may improve the agreement, because
LTE models systematically overestimate radiative cooling and thus underestimate the temperature.
Chemiluminescence by H$_2$O, formed by the exothermic reaction of OH+H$_2$, might also be expected. 

Although our computed HS and S$_2$ column densities increase with metallicity,
 optically thick columns are predicted for all plausible atmospheric compositions,
which means that millibar-level temperature inversions are expected to be commonplace.
The distinctive interaction of S$_2$ and HS with near ultraviolet light
could make these species detectable in transit by the refurbished HST; 
there is evidence for a blue absorber in legacy HST data of HD 209458b (Sing et al 2008).

On the other hand, sulfur does not give an easy answer to why some hot Jupiters have superheated
stratospheres, and others not.  
In an earlier draft of this study, we speculated that
S$_3$---which is very sensitive to metallicity---might be part of the explanation.
This no longer appears likely.
We have since developed a better understanding of HS's opacity, which turns out to be considerable.
We no longer see a strong connection between metallicity and radiative heating, save at very low pressures ($<\!100 \mu$bars)
where SO and SO$_2$ become important.
It now seems that sulfur chemistry by itself is unlikely to explain differences between planets,
although planetary metallicity may still be key. 

Heating by sulfur compounds does not preclude heating by TiO and VO on hotter planets.
Sulfur species provide considerable heating from below 1000 K to above 2000 K, but they do not provide the spectral coverage
at visible wavelengths that TiO and VO provide.
For TiO and VO to be abundant enough to explain stratospheric heating, the temperature needs to be very high, in excess of 2000 K,
and not just in the stratosphere but also at deeper levels in the planet where these
two refractory oxides would otherwise be cold-trapped in silicate clouds.
OGLE-TR-56b (Sing and L{\'o}pez-Morales 2009) seems to meet the TiO-VO threshold.

CO$_2$ is generated by the reaction of CO with OH and destroyed by the reverse (endothermic) reaction with H, CO + OH $\leftrightarrow$ CO$_2$ + H.  
At low altitudes OH is generated by the reaction of H$_2$O with atomic H, supplemented at high altitudes by UV photolysis of H$_2$O.
As both the major source and major sink of CO$_2$ are proportional to atomic hydrogen densities,
the kinetic inhibition against hydrogen recombination does not disturb CO$_2$'s thermochemical equilibrium.
We find that CO$_2$ mixing ratios vary quadratically with metallicity from $1.6\times10^{-8}$ to
$1.6\times 10^{-4}$ for $0<[{\rm M}/{\rm H}]<0.7$.
This result is insensitive to insolation, vertical mixing, temperature (for $1200 \leq T \leq 2000$ K), and gravity.
Because the reactions that form and destroy CO$_2$ leave the total number of molecules unchanged,
the CO$_2$/CO ratio is also pressure independent. 
The computed CO$_2$ abundances are in good agreement with the
observation of CO$_2$ at the ppmv level in HD 189733b (Swain et al 2009).
Therefore we confirm Lodders and Fegley's (2002) suggestion that CO$_2$ is a promising probe of planetary metallicity.

\section{Acknowledgements}
We thank R.\ V.\ Yelle for discussions regarding the potential importance of S$_3$,
and G.\ Tinetti for an insightful review.
We thank NASA's Exobiology and Planetary Atmospheres Programs for support. 
KL was also supported by NSF Grant AST-0707377.

\begin{table}[htdp]
%\caption{\small }
{\footnotesize
%{\scriptsize
\begin{tabular}{lcrcccrrrrr} % 11
\hline
Model &  $[{\rm M}/{\rm H}]^a$ & $I^b$ & $T$ & S$_2$ [cm$^{-2}$]$^c$ & HS [cm$^{-2}$]$^c$ & HS$^d$ & SO$^d$ & CO$_2^d$ & Heating$^e$ & Heating$^f$\\
\hline
A & $0$ & 1000 & 1400 &  $4.2\times 10^{18}$ & $1.2\times 10^{20}$ & 6 & $0.07$  & $0.065$ & $6.4\times 10^{4}$ & $2.7\times 10^{4}$\\
M & $0.7$ & 1000 & 1400 &  $2.2\times 10^{20}$ & $1.0\times 10^{21}$ & 26  & $1.7$  & $1.6$ & $8.4\times 10^{4}$ & $4.1\times 10^{4}$\\
MM & $1.4$ & 1000 & 1400 &  $6.0\times 10^{21}$ &$5.4\times 10^{21}$ & 100 & $27$  & $41$ & $1.1\times 10^{5}$& $5.1\times 10^{4}$\\
H & $0.7$ & 1000 & 1600 &  $2.2\times 10^{20}$ &$2.3\times 10^{21}$  & 43 & $1.2$  & $1.4$ & $9.0\times 10^{4}$& $4.3\times 10^{4}$\\
HH & $0.7$ & 1000 & 1800 &  $1.9\times 10^{20}$ &$4.0\times 10^{21}$  & 52 &  $1.3$  & $1.5$& $9.5\times 10^{4}$ & $4.4\times 10^{4}$\\
HHH & $0.7$ & 1000 & 2000 &  $1.3\times 10^{20}$ &$5.0\times 10^{21}$ & 41 & $1.4$  & $1.3$ & $9.6\times 10^{4}$ & $4.1\times 10^{4}$\\
C  & $0.7$ & 1000 & 1200 &  $1.6\times 10^{20}$ &$2.5\times 10^{20}$  & 11 & $1.9$  & $1.9$ & $7.5\times 10^{4}$& $3.7\times 10^{4}$\\
G & $0.7$ & 1000 & 1400 &  $4.3\times 10^{20}$ & $2.0\times 10^{21}$  & 32 & $1.3$  & $1.6$ & $9.3\times 10^{4}$ & $4.7\times 10^{4}$\\
I & $0.7$ & 200 & 1400 &  $2.1\times 10^{20}$ & $1.0\times 10^{21}$  & 37 & $0.9$  & $1.6$ & $1.7\times 10^{4}$& $8.8\times 10^{3}$\\
SSC  & $0.7$ & 1 & 1200 &  $1.1\times 10^{20}$ &$2.4\times 10^{20}$ & 16 & $0.06$  & $1.9$ & $72$ & $44$\\
%X$^g$ &  $10^{7}$ & $1$ & 1000  & 20 & 1400 &  $3.3\times 10^{18}$ & $8.2\times 10^{18}$  & $2.2\times 10^{12}$  & $0.05$ & 1.7 & $1.2\times 10^{4}$\\
\hline
\multicolumn{11}{l}{$a$ -- Metallicity.  This notation means that the planet is $10^{[{\rm M}/{\rm H}]}$ richer in C, S, N, and O than the Sun.}\\
\multicolumn{11}{l}{$b$ -- Insolation.  $I=1000$ corresponds to a solar twin primary at 0.032 AU.}\\
\multicolumn{11}{l}{$c$ -- Column densities above 1 bar.}\\
\multicolumn{11}{l}{$d$ -- Mixing ratio in ppmv at 1 mbar.}\\
\multicolumn{11}{l}{$e$ -- Total atmospheric heating [W/m$^2$] above 0.1 bar for a solar twin source.}\\
\multicolumn{11}{l}{$f$ -- Total atmospheric heating [W/m$^2$] above 1 mbar for a solar twin source.}\\
%\multicolumn{12}{l}{$g$ -- Model X uses a different, obsolete list of reactions (Zahnle et al 1995).}\\
\hline
\end{tabular}
}
\label{table_one}

\end{table}

%\begin{figure}[!htb] %  figure placement: here, top, bottom
%   \centering
%\includegraphics[width=1.0\textwidth]{Model_A_16.pdf} % requires the graphicx package
%   \caption{\small Important sulfur species, CO, and CO$_2$ in the atmosphere of a hot Jupiter with solar metallicity.  
%The atmosphere is assumed isothermal at 1400 K and insolated 1000$\times$ more strongly than Earth.  
%Other model A parameters are listed in Table 1.  The prominent transition at $\sim$2 mbar --- the altitude where the S$_2$ mixing ratio peaks --- is associated with photolysis of H$_2$S.  Abundances in the 1400 K atmosphere are generally representative of atmospheres with $1200<T<2000$ K.  }
%\label{fig:model_A}
%\end{figure}

\newpage
\small


\begin{thebibliography}{31}


\item {Allard}, F., {Hauschildt}, P.~H., {Alexander}, D.~R.,
{Tamanai}, A., \&
   {Schweitzer}, A. 2001, \apj, 556, 357

\item {Barman}, T.~S. 2008, \apjl, 676, L61

\item {Barnes}, J.~R., {Barman}, T.~S., {Jones}, H.~R.~A., {Leigh},
C.~J., {Cameron},
   A.~C., {Barber}, R.~J., \& {Pinfield}, D.~J. 2008, \mnras, 390, 1258
   
 \item Berdyugina, S.V. and Livingston, W.C. (2002) {\em Astron.\ Astrophys.\ 387,} L6-L9. 

\item Billmers RI and Smith AL (1991).  {\em J.\ Phys.\ Chem.\ 95,}
4242-4245.

\bibitem[{{Burrows} {et~al.}(2007){Burrows}, {Hubeny}, {Budaj}, {Knutson}, \&
  {Charbonneau}}]{Burrows07c}
{Burrows}, A., {Hubeny}, I., {Budaj}, J., {Knutson}, H.~A., \& {Charbonneau},
  D. 2007, \apjl, 668, L171

\bibitem[{{Burrows} {et~al.}(2008){Burrows}, {Budaj}, \& {Hubeny}}]{Burrows08}
{Burrows}, A., {Budaj}, J., \& {Hubeny}, I. 2008, \apj, 678, 1436

\bibitem[{{Burrows} {et~al.}(2008){Burrows}, {Budaj}, \& {Hubeny}}]{Burrows08}
{Burrows}, A., {Budaj}, J., \& {Hubeny}, I. 2008, \apj, 678, 1436

\bibitem[{{Charbonneau} {et~al.}(2008){Charbonneau}, {Knutson}, {Barman},
  {Allen}, {Mayor}, {Megeath}, {Queloz}, \& {Udry}}]{Charb08}
{Charbonneau}, D., {Knutson}, H.~A., {Barman}, T., {Allen}, L.~E., {Mayor}, M.,
  {Megeath}, S.~T., {Queloz}, D., \& {Udry}, S. 2008, \apj, 686, 1341

\item Claire MW, Catling DC, Zahnle KJ (2006).  {\em Geobiology} {\bf 4,}  239-269.

\item Dalgarno A and Williams DA (1962). .
{\em Astrophy.\ J.\ 136,} 690-692.

\item {D{\'e}sert}, J.-M., {Vidal-Madjar}, A., {Lecavelier Des Etangs}, A., {Sing}, D., {Ehrenreich}, D., {H{\'e}brard}, G., \& {Ferlet}, R. 2008,
\aap, 492, 585

\bibitem[{{Fegley} \& {Lodders}(1994)}]{Fegley94}
{Fegley}, B.~J. \& {Lodders}, K. 1994, Icarus, 110, 117

\bibitem[{{Fortney} {et~al.}(2006){Fortney}, {Saumon}, {Marley}, {Lodders}, \&
  {Freedman}}]{Fortney06}
{Fortney}, J.~J., {Saumon}, D., {Marley}, M.~S., {Lodders}, K., \& {Freedman},
  R.~S. 2006, \apj, 642, 495

\bibitem[{{Fortney} {et~al.}(2008){Fortney}, {Lodders}, {Marley}, \&
  {Freedman}}]{Fortney08a}
{Fortney}, J.~J., {Lodders}, K., {Marley}, M.~S., \& {Freedman}, R.~S. 2008,
  \apj, 678, 1419

\bibitem[{{Harrington} {et~al.}(2007){Harrington}, {Luszcz}, {Seager},
  {Deming}, \& {Richardson}}]{Harrington07}
{Harrington}, J., {Luszcz}, S., {Seager}, S., {Deming}, D., \& {Richardson},
  L.~J. 2007, \nat, 447, 691

\item Henneker, W. H. and Popkie, H.E. (1971) {\em J.\ Chem.\ Phys.\  54,} 1763-1778.

\bibitem[{{Hubeny} {et~al.}(2003){Hubeny}, {Burrows}, \& {Sudarsky}}]{Hubeny03}
{Hubeny}, I., {Burrows}, A., \& {Sudarsky}, D. 2003, \apj, 594, 1011

\item Kasting JF, Zahnle KJ, Pinto JP, Young AT (1989)   {\em Origins of Life,} \textbf{19}, 95-108.

\item Kasting JF (1990). {\em Origins of Life  20,} 199-231.

\bibitem[Knutson et al.(2008)]{2008ApJ...673..526K} Knutson, H.~A., 
Charbonneau, D., Allen, L.~E., Burrows, A., 
\& Megeath, S.~T.\ 2008, \apj, 673, 526 

\bibitem[Knutson et al.(2009)]{2009ApJ...691..866K} Knutson, H.~A., 
Charbonneau, D., Burrows, A., O'Donovan, F.~T., 
\& Mandushev, G.\ 2009, \apj, 691, 866 


\item  Larsson, M. (1983) {\em Astron.\ Astrophys.\ 128,} 291-298.

\item Liang M-C, Parkinson CD, Lee AYT, Yung YL, and Seager S, (2003) ``Source of atomic hydrogen in the atmosphere of HD 209458b'' {\em Astrophys.\ J.\ 596,} L247ÐL250.

\item Liang M-C, Seager S, Parkinson CD, Lee AYT, and Yung YL (2004) ``On the insignificance of photochemical hydrocarbon aerosols in the atmospheres of close-in extrasolar giant planets'' {\em Astrophys.\ J.\ 605,} L61ÐL64.


\item Lodders, K. and Fegley B. (1999) {\em The Planetary Scientist's Companion.} Oxford.

\bibitem[{{Lodders}(1999)}]{Lodders99}
{Lodders}, K. 1999, \apj, 519, 793

\bibitem[{{Lodders}(2002)}]{Lodders02b}
---. 2002, \apj, 577, 974

\item Lodders, K.\ and Fegley, B.J.\ (2002). {\em Icarus 155,} 393-424.

\item Machalek P, McCullough PR, Burke CJ, Valenti JA, Burrows A, and Hora JL (2008) {\em Astrophys.\ J.\ 684,} 1427-1432.

\bibitem[{{Marley} {et~al.}(2007){Marley}, {Fortney}, {Seager}, \&
  {Barman}}]{Marley07b}
{Marley}, M.~S., {Fortney}, J., {Seager}, S., \& {Barman}, T. 2007, in
  Protostars and Planets V, ed. B.~{Reipurth}, D.~{Jewitt}, \& K.~{Keil},
  733--747

\item Moses JI, Zolotov MY, and Fegley B (2002). {\em Icarus 156,} 76Ð106.

\item Nicholas, J.E., Amodio, C.A., Baker, M.J. (1979)   {\em J.\ Chem.
\ Soc.\ Faraday Trans.\ 1:75, }1868-1880.

\item Noll KS, McGrath MA, Trafton LM, Atreya SK, Caldwell JJ, Weaver HA, Yelle RV, Barnet C, and Edgington S (1995) {\em Science, 267,} 1307.

\item Okabe H (1978) {\em The Photochemistry of Small Molecules.}
Wiley-Interscience, New York, 431 pp.

%\item  Pen, J, Hu, X, Marshall, P. (1999)    {\em J.\ Phys.\ Chem.\ A 103,} 5307 - 5311.

\item Ram, R S, Bernath P. F., Engleman R. and Brault J. W. (1995) {\em J.\ Molec.\ Spect. 172,} 34-42.

\item Resende, S.M., and Ornellas, F.R. (2001) {\em J.\ Chem.\ Phys.\ 115,} 2178-2187.

\bibitem[{{Richardson} {et~al.}(2007){Richardson}, {Deming}, {Horning},
  {Seager}, \& {Harrington}}]{Richardson07}
{Richardson}, L.~J., {Deming}, D., {Horning}, K., {Seager}, S., \&
  {Harrington}, J. 2007, \nat, 445, 892

%\item Sander SP, Friedl RR,  Ravishankara AR, Golden DM, Kolb CE, Kurylo MJ, Huie RE, Orkin VL, Molina MJ, Moortgat GK, and Finlayson-Pitts BJ (2003)
%Chemical Kinetics and Photochemical Data for Use in Atmospheric Studies. Evaluation Number 14. JPL Publication 02-25.

\item Sauval, A.J. and Tatum, J.B. (1984) {\em Astrophs.\ J.\ Supp.\ 56,} 193-209.

\item Schneider, L., Meier W, and Weige KH (1990) {\em J.\ Chem.\ Phys.\ 92,} 7027-7037.

%\item Shiina, H., Oya, M., Yamashita, K., Miyoshi, A., and Matsui, H. (1996)   {\em Phys.\ Chem.\ 100,} 2136-2140.

\bibitem[{{Showman} {et~al.}(2008){Showman}, {Fortney}, {Lian}, {Marley},
  {Freedman}, {Knutson}, \& {Charbonneau}}]{Showman09}
{Showman}, A.~P., {Fortney}, J.~J., {Lian}, Y., {Marley}, M.~S., {Freedman},
  R.~S., {Knutson}, H.~A., \& {Charbonneau}, D. 2008, ArXiv e-prints
  
\item Sing, D.K., Vidal-Madjar A., D{\'e}sert, J.-M., Lecavelier des Etangs, A., and Ballester, G.\ (2008). {\em Astrophys.\ J.\ 686,} 658-666.

\item Sing, D.K., and L{\'o}pez-Morales, M. (2009). {\em Astron.\ Astrophys.\ 493,} L31-L34.

\item Spencer JR., Jessup, KL, McGrath MA, Ballester GE, amd Yelle RV (2000). {\em Science 288,} 1208-1210.


\item Steudel R and Eckert B (2003). {\em Elemental sulfur and sulfur-rich compounds.}  Springer, 202 pp.

\item Swain, M.~R., Vasisht, G., Tinetti, G., Bouwman, J., Chen, P.,
Yung, Y., Deming, D.,
\& Deroo, P.\ 2009, {\em Astrophys.\ J.\ Lett.\ 690,} L114-L117.

%\item Tesner, P.A., Nemirovskii, M.S., and Motyl, D.N. (1990 )   {\em Kinet.\ Catal.\ 31,} 1081-1083.

\item Tinetti, G.\ (2008).  {\em Bull.\ Am.\ Astron.\ Soc.\ 40,} 463.

\bibitem[{{Torres} {et~al.}(2008){Torres}, {Winn}, \& {Holman}}]{Torres08}
{Torres}, G., {Winn}, J.~N., \& {Holman}, M.~J. 2008, \apj, 677, 1324

\item van der Heijden and van der Mullen (2001). {\em J.\ Phys.\ B.
Atom.\ Mol.\ Opt.\ Phys.\ 34,} 4183-4201.

\item Visscher C, Lodders K, and Fegley B. (2006). {\em Astrophys.\ J.\ 648,} 1181-1195.

%\item Yelle RV, and McGrath MA (1996). {\em Icarus 119,} 90-111.

\item Wheeler, M.D., Orr-Ewing, AJ, and Ashfold, MNR (1997) {\em J.\ Chem.\ Phys.\ 107,} 7591-7600. 

\item Zahnle KJ, Mac Low M-M, Lodders K, and Fegly B. (1995). {\em Geophys.\ Res.\ Lett.\ 22,} 1593-1596.

\item Zahnle KJ, Claire MW, Catling DC (2006).   {\em Geobiology 4,} 271-282.

\item Zahnle KJ, Haberle RM, Catling DC, Kasting JF (2008).  {\em J.\ Geophys.\ Res.\ 113,}  E11004, doi:10.1029/2008JE003160.

\item Zare, R.N., Schmeltekopf AL, Harrop WJ, and Albritton DL (1973) {\em J.\ Molec.\ Spect.\ 46,} 37-66.

\end{thebibliography}
\end{document}